\documentclass[review,sort&compress]{elsarticle}
\usepackage{url,lineno}
\usepackage[utf8]{inputenc}
\usepackage{makecell}
\usepackage{graphics}
\usepackage{xspace}

\usepackage{booktabs} 
\usepackage{amsfonts, amsmath, amsfonts}
\usepackage{textcomp}
\usepackage[english]{babel}
\usepackage{blindtext}
\usepackage{xcolor}
\usepackage{colortbl}
\usepackage[linesnumbered,algo2e,ruled]{algorithm2e}
\usepackage{tabularx}
\usepackage{algorithm}
\usepackage{listings}
\usepackage[thinlines]{easytable}
\usepackage{algpseudocode}
\usepackage{svg}
\usepackage{bm}
\usepackage{pifont}
\usepackage{tablefootnote}
\usepackage{longtable}
\usepackage{multirow}
\usepackage{booktabs}
\usepackage[hidelinks, bookmarks=false]{hyperref}
\usepackage{siunitx}
\usepackage{amsthm}
\usepackage{booktabs,subcaption,dcolumn}
\usepackage{soul}
\usepackage{mathtools, nccmath}
\usepackage{flushend}
\usepackage[export]{adjustbox}[2011/08/13]

\usepackage{tikz}

\newcommand\revision[1]{%
  \bgroup
  \hskip0pt\color{blue!80!black}%
  #1%
  \egroup
}

\newcommand\delete[1]{%
  \bgroup
  \hskip0pt\color{red!80!black}%
  #1%
  \egroup
}

\newcommand{\codeword}[1]{\texttt{#1}}
\newcommand{\file}[1]{\textit{#1}}
\newcommand\MLalgorithm[1]{{\fontfamily{qhv}\selectfont #1}}

\journal{Computers \& Security}

\begin{document}

\begin{frontmatter}

\title{Classification of Web Phishing Kits for early detection by platform providers}

\author[1]{Andrea Venturi\corref{cor1}}
\ead{andrea.venturi@unimore.it}

\author[2]{Michele Colajanni}
\ead{michele.colajanni@unibo.it}

\author[3]{Marco Ramilli}
\ead{marco.ramilli@yoroi.company}

\author[1]{Giorgio Valenziano Santangelo}
\ead{giorgio.valenziano@unimore.it}

\cortext[cor1]{Corresponding author}

\address[1]{
    Department of Engineering ``Enzo Ferrari'', University of Modena and Reggio Emilia, Italy
}

\address[2]{
    Department of Computer Science and Engineering, University of Bologna, Italy
}

\address[3]{
    Yoroi, Italy
}

\begin{abstract}
Phishing kits are tools that dark side experts provide to the community of criminal phishers to facilitate the construction of malicious Web sites. As these kits evolve in sophistication, providers of Web-based services need to keep pace with continuous complexity.
We present an original classification of a corpus of over 2000 recent phishing kits according to their adopted evasion and obfuscation functions. We carry out an initial deterministic analysis of the source code of the kits to extract the most discriminant features and information about their principal authors. We then integrate this initial classification through supervised machine learning models. Thanks to the ground-truth achieved in the first step, we can demonstrate whether and which machine learning models are able to suitably classify even the kits adopting novel evasion and obfuscation techniques that were unseen during the training phase. 
We compare different algorithms and evaluate their robustness in the realistic case in which only a small number of phishing kits are available for training. 
This paper represents an initial but important step to support Web service providers and analysts in improving early detection mechanisms and intelligence operations for the phishing kits that might be installed on their platforms.

\end{abstract}

\begin{keyword}
    Phishing kits \sep Machine learning \sep Evasion \sep Obfuscation \sep Early detection
\end{keyword}

\end{frontmatter}


\section{Introduction}
\label{sec:introduction}

Phishing is an evergreen cyber attack leveraging social engineering methods.
From the initial generic and na\"{i}ve attacks, nowadays phishing is evolving in precision and sophistication with the historical goal of gathering credentials of unaware users~\cite{aleroud2017phishing}. Due to its efficacy, phishing methods are also being used as vectors to spread malware and as a first step to perform other malicious operations~\cite{chiew2018survey}.
From the point of view of an attacker, the creation of a counterfeit copy of a legitimate website and its deployment are laborious and time-consuming phases even because they have to change frequently. \textit{Phishing kits} are created to automate these cumbersome actions for the phishers. Facilitating the implementation and exploitation of phishing websites expands the number of attackers to even criminals with limited experience to the extent that now they represent a serious menace to cybersecurity~\cite{cova2008there, bijmans2021catching}.

Previous research results analyze specific phishing kits and peculiar functions, but supporting the provider detection of this type of attack deserves a deeper and broader study.
We present the first classification that groups phishing kits according to their adopted evasion and obfuscation functions. This analysis represents an opportunity to understand some technical features of the phishing business, and it reveals fundamental information about malicious techniques that security experts and hosting providers can use to improve authorship attribution and detection mechanisms.

Any classification of phishing kits is challenging because the authors continuously modify the previous versions by introducing novel evasion and obfuscation techniques that complicate the analysis and detection. 
However, we observe that the most prolific authors introduce new functions but tend to reuse (many) parts of existing material. From a defensive point of view, a similar redundancy represents a key factor because we can provide defenders with a classification of phishing kits based on the most common malicious functions and characteristics.
Our proposal consists of two main phases. 

The first phase involves static analysis techniques carried out on the source code of over 2000 recent phishing kits. Our goal is to perform a deterministic classification that groups the kits according to the most frequent evasion and obfuscation techniques that are adopted by the authors. This classification points out common patterns in the habits of the most popular authors of phishing kits. We should consider that, as phishing is now a mature industry where a small community of expert authors design and implement the phishing kits, it is reasonable to expect that there is a correlation between each author and his habits in employing specific evasion and obfuscation functions.
We confirm that the most prolific authors tend to maintain similar development habits through time. This result can support threat intelligence operations in author profiling operations. 

Related literature usually adopts a similar deterministic approach when dealing with phishing kits~\cite{merlo2022phishing, oest2018inside}. However, any static analysis on past kits cannot help to detect unknown techniques. As the evolution of phishing kits is continuous, existing deterministic approaches represent a serious limit. For this reason, we propose an additional probabilistic classification based on specific Machine Learning (ML) models. The idea is to train supervised ML models to recognize evasive and obfuscated kits from structural information of the phishing kits and authors' design patterns. 
Through a large set of experiments, we investigate which ML models are the most suitable in which conditions. We can conclude that most ML classifiers guarantee good performance in the identification of evasive and obfuscated kits. Our approach leads to a probabilistic classification that can flag even the kits that employ novel evasion and obfuscation techniques that were not considered during training thus addressing the main issues of existing approaches. 

An interesting and initially unexpected outcome is that the results of the majority of the proposed classifiers are almost insensitive to the dimension of the dataset used in the training phase. More precisely, even training on just 20\% of the available phishing kits does not worsen much the results with respect to training experiments carried out on the typical 80\% of the dataset, despite a slight drop in the classification precision. 
Although we evaluate our method on a vast corpus of phishing kits, they represent a small fraction of existing and novel kits. Hence, the robustness of our results trained on small sets demonstrates the effectiveness of our proposal in classifying even future new kits. These results can be exploited by host and service providers to boost early detection mechanisms by keeping track of the most employed evasion and obfuscation functions tailored to prevent an easy identification of the phishing kits.

The remainder of this paper is organized as follows. Section \ref{sec:background_bis} contains some background information about phishing kits, their business and how our analysis can be useful to security practitioners.
Section \ref{sec:classification} describes the phases of our method. Section \ref{sec:deterministic} presents the results of the deterministic classification, while Section \ref{sec:ml_results} details machine learning algorithms and considered scenarios. Section~\ref{sec:ml_evaluation} discusses the results of the probabilistic classification. Section \ref{sec:related} compares our work against related literature. Finally, Section \ref{sec:conclusions} draws the conclusions and presents some possible extensions of the results of this paper.

\section{Phishing kits}
\label{sec:background_bis}

\subsection{Goal}

Phishing is designed to fool unwary victims by making them believe that they are interacting with a legitimate website, and inducing them to insert their credentials~\cite{das2019sok}. The popularity of this attack continues to grow, with thousands of new phishing URLs found every day~\cite{apwg2022report}.

Phishing attacks involve several phases. The attackers need to initially implement a website mimicking that of the targeted organization. This site must be integrated with web forms to collect user credentials or other information. The malicious website needs also to be deployed on an HTTP server in order to be accessible to the victims. To lure a large number of users, the attackers usually choose a deceiving URL for the phishing website reminding the one of the legitimate page. Only afterward, they begin the phishing message campaign aimed to attract the victims to the malicious page and induce them to insert their credentials~\cite{alabdan2020phishing}.

Among all these operations, the implementation of a counterfeit copy of a legitimate website is certainly the most time-consuming and error-prone operation. This element and the fact that phishing campaigns span just a few hours (as indicated by last estimations~\cite{oest2020sunrise}) have pushed expert phishers to implement \textit{phishing kits} that automate this procedure. These kits consist of archives containing all the necessary components to deploy a fake website~\cite{bijmans2021catching}. The initial versions of phishing kits were simple HTML pages with minimal back-end source code (usually written in PHP) used to log the stolen credentials or eventually to send them to the phisher. However, modern versions consist of complex software functions that include credit card skimming, malware spreading and obfuscation operations. 
In practice, the phishers in possession of a phishing kit can proceed as follows to obtain a complete working environment: first, they acquire an operational web server or select a public one; then, they upload the phishing kit and extract its content on it; finally, they start sending the luring messages. This set of facilities is behind the continuous growth of phishing attacks in recent years.

\subsection{Detection}
\label{sec:early_det}

From a defensive point of view, phishing detection mechanisms operate at multiple levels. Extensive work is aimed to detect phishing pages directly from the deployment URLs or from the aspect of the HTML page. At the same time, numerous tools recognize phishing emails analyzing their content. Despite their efficacy, all these methods suppose that the phishing web-page is already deployed. Other solutions pursue an early detection, in which the same web-server providers apply filtering mechanisms to exclude the malicious code as soon as it is uploaded to the server. While the anti-phishing mechanisms continue to improve, phishing kits are evolving as well, with modern versions implementing several evasion and obfuscation techniques that prevent or delay detection~\cite{zhang2021crawlphish}. 
Evasion functions aim to elude the detection of defensive tools by forbidding access to unwanted visitors, excluding the traffic from cybersecurity companies and automatic search engines or employing complex redirection mechanisms~\cite{oest2019phishfarm, han2016phisheye}. On the other hand, obfuscation techniques are used to conceal the original source code, thus complicating static analyses and reverse engineering investigations~\cite{you2010malware, cova2008there, proofpoint2016obfuscation}. 

Web-server providers need to keep pace with these functions to devise updated early detection mechanisms. Our study pursues this goal, as it provides a novel classification of phishing kits according to the evasion and obfuscation functions implemented in their source code. Our innovative approach is meant not only to detect current evasion and obfuscation methods employed by modern kits, but also to predict the usage of unknown methods through machine learning algorithms. Our results can be exploited by service providers to understand which are the most frequent functions, and to prioritize mitigating countermeasures.

\subsection{Phishing kit business}
\label{sec:pk_businees_bis}

\begin{figure*}[t!]
    \centering
    \includegraphics[width=1.3\columnwidth, center]{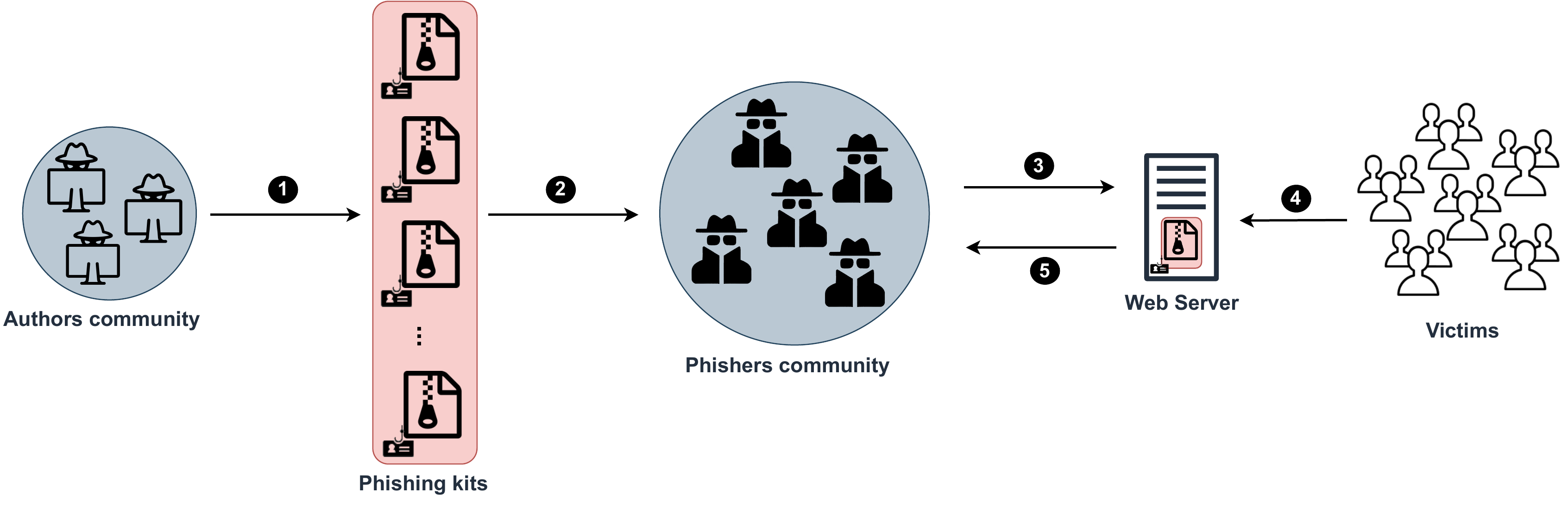}
    \caption{Phishing-as-a-Service workflow.}
    \label{fig:phaas-bis}
\end{figure*}

Cybercrime is now a mature industry that is characterized by specializations together with markets for the exchange and sale of products. Even phishing has reached the maturity of the \textit{Crime-as-a-Service} business model where expert cybercriminals sell or rent malicious products and services to third-party actors that do not necessarily have strong technical skills. Nowadays, phishing kits represent a product of the \textit{Phishing-as-a-Service} business~\cite{chiew2018survey}. In Figure~\ref{fig:phaas-bis} we outline the typical workflow where an expert author implements a phishing kit (step 1) and provides it to the phishers community through different business relationships (step 2). Each kit typically comes with a dedicated section in which a user can specify his preferred email address(es) where to receive stolen credentials. Once a phisher has bought a kit, it is sufficient that he adds his destination lists and uploads the kit on a controlled web server. He then extracts its content and begins the phishing email campaign (step 3). Finally, as a typical fisherman, he waits for victims to interact with the uploaded website (step 4) and collects the exfiltrated information (step 5).

For the goal of this paper, it is important to observe that the authors of the phishing kits and their customers (phishers) collaborate, but they typically belong to different groups. The community of expert authors is much smaller than that of the phishers because this business model facilitates even inexpert criminals to carry out sophisticated phishing attacks. 

As in any business model, the authors want to maximize their profits. This can be achieved through higher prices or by selling more products. The latter choice is more frequent, hence it can be expected that the phishing kit authors tend to re-use (even large) parts of their previous material, as confirmed by us and related works (e.g.,~\cite{bijmans2021catching, oest2018inside, merlo2022phishing}).

We leverage this hypothesis for the classification of the over 2000 analyzed phishing kits. Moreover, such re-utilization allows us to highlight common coding habits of the principal phishing kit authors. As many of them intentionally leave their signature inside the kits, we devise author profiles that map the most productive ones to the evasion and obfuscation functions considered in our method. Hence, besides the benefits related to detection, our classification can support threat intelligence operations which aim to track through time the coding behaviors of the principal authors and to catch possible evolutionary changes.     



\section{Classification method}
\label{sec:classification}

Phishing kits are growing in number, sophistication and complexity~\cite{aleroud2017phishing}. As a consequence, the manual analysis of cybersecurity experts is becoming more complicated. We propose a novel classification method aimed to group phishing kits according to the offered evasion and obfuscation functions. Our goal is to design a method capable of keeping pace with the continuous introduction of new techniques in modern kits and to better understand malicious habits of the principal authors.

\begin{figure*}[t!]
    \centering
    \includegraphics[width=1.4\columnwidth, center]{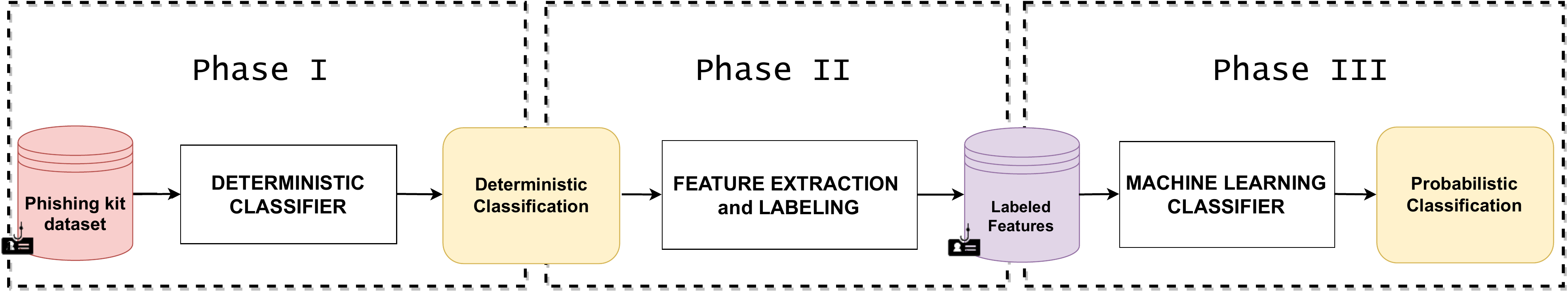}
    \caption{The three phases of the proposed classification method.}
    \label{fig:automatic_classification}
\end{figure*}

The overall methodology consists of the three main phases that are shown in Figure \ref{fig:automatic_classification}. Phase I involves a static analysis of the content of the phishing kits and produces a deterministic classification on the basis of the employed evasion and obfuscation functions. This analysis is able to recognize known techniques, but it cannot detect new versions or novel functions that phishing kit authors might introduce. Phase II aims to extract numerous features which summarize the structure and the functions of each kit along with the authors' design patterns, and to label them according to the deterministic results. Phase III exploits all previous information and obtains a probabilistic classification through ML algorithms. This last classification aims to detect evasive and obfuscated kits on the basis of structural information of the phishing kits and authors' design patterns. This contribution is important because it allows the ML classifiers to overcome the typical drawbacks of the deterministic approaches (similar to that in Phase I) that cannot chase the authors of phishing kits when they introduce new evasion and obfuscation techniques or modify existing ones. 
This result is possible because the authors of the prevalent phishing kits belong to a small community, and the most prolific of them augment productivity by re-using parts of existing kits~\cite{bijmans2021catching, oest2018inside, cui2017tracking}. 


The contributions of the three phases are detailed below.

\subsection{Phase I - Deterministic classification}
\label{sec:phaseI}
The first phase involves several operations that statically analyze the phishing kit structure and its source code with the goal of classifying phishing kits according to the employed evasion and obfuscation techniques.
Static analysis extracts behavioral patterns of software without executing it~\cite{gandotra2014malware}. Similarly, our approach parses the source code of the kits, and leverages heuristics and regular expressions to determine whether the kit employs known evasion or obfuscation functions.
Despite the limited number of false positives granted by static analysis, any deterministic approach is unable to detect novel evasion and obfuscation techniques and minor modifications of existing functions~\cite{ucci2019survey}.

\subsection{Phase II - Feature extraction and labeling}
\label{sec:phaseII}
An exclusively deterministic approach is insufficient to classify those kits that employ unknown evasion and obfuscation techniques. For this reason, the second phase of our method extracts multiple features and authors' design patterns from the phishing kits, and labels them in preparation for the third phase.
Although many authors introduce novel functions, they tend to re-use parts of existing kits~\cite{oest2018inside, cui2017tracking}. Moreover, due to the growing popularity of the Phishing-as-a-Service business model in the criminal ecosystem~\cite{alabdan2020phishing}, only a small group of expert authors designs, implements and sells entirely new phishing kits to the phishers community~\cite{bijmans2021catching}. Because of this re-utilization, many characteristics of various phishing kits are preserved. This motivates our choice of modeling authors' design patterns and stylistic choices into our feature set. As it is unlikely that authors change their habits even when introducing new techniques~\cite{kalgutkar2019code}, in the third phase we will leverage these features to obtain the final probabilistic classification through machine learning classifiers.

Each kit is then labeled as follows:
\begin{itemize}
    \item A kit is flagged as \textbf{Evasive} if the deterministic classification of Phase I detects at least one of the considered evasion techniques.
    \item A kit is flagged as \textbf{Obfuscated} if the deterministic classification of Phase I detects at least one of the considered obfuscation techniques.
\end{itemize}
We refer the reader to the next section for a list of the considered features, evasion and obfuscation techniques.

At the end of the second phase, we obtain a labeled dataset that can be submitted to the ML classifiers of the third phase.

\subsection{Phase III - Probabilistic classification}
\label{sec:phaseIII}

Phase III aims to achieve the probabilistic classification which let us overcome the limitations of a deterministic approach. Phase II produces a representation of each phishing kit in terms of its structural features and authors' design pattern choices. These kits have also been labeled according to the evasion and obfuscation techniques that were detected in Phase I. In this way, we are able to train supervised ML classifiers that can predict evasive and obfuscated kits from their features. We leverage the generalization ability of ML classifiers because phishing kits tend to continuously evolve. Hence it is fundamental to have classification methods that are able to adapt their results to new functions.
After they have learned the structural choices and design patterns of evasive and obfuscated kits, we expect that the resulting classifiers are able to detect the usage of known and also novel techniques.

Let us explain it with an example. Let us assume that a new phishing kit needs to be analyzed and classified, and that the same author has produced other kits already labeled as evasive because Phase I detected a specific evasion function in all of them. As it is a novel kit, it contains a new evasion technique that is not recognized by the deterministic classification. However, as the author tends not to change his developing habits, it is reasonable to expect that he re-utilizes portions of his previous kits. This behavior is captured in our feature set, and the resulting classifiers can mark the novel kit as evasive with high confidence. This indication can be exploited by cybersecurity defenders (e.g., maintainers of datacenters that are likely to host phishing kits) to identify the novel function, and consequently, design a countermeasure that will improve existing detection tools. Moreover, security experts can also build authors' profiles for advanced threat intelligence operations.
\section{Deterministic classification of phishing kits}
\label{sec:deterministic}

The deterministic analysis of Phase I is applied to the dataset of the phishing kits that are contained in Phishing Kits Tracker~\cite{ramilli2020tracker}. Phishing Kits Tracker consists of a publicly available collection of phishing kits that are harvested from malicious websites. To the best of our knowledge, this is the first research work that employs a similar dataset.
Phishing Kits Tracker is daily updated by an automatic back-end script maintained by one of the authors of this paper, and contains the most recent phishing kits. At the time of the analysis described in this paper, it counts $2245$ phishing kits. Its size makes our work one of the most comprehensive and detailed classification of phishing kits that has ever been published. Each kit in the dataset is reported in its raw version with no pre-processing pruning operation. For this reason, the considered collection contains a large variety of data: from small kits consisting of just basic files to deploy a phishing website, to large sophisticated kits of hundreds of megabytes.

\subsection{Evasion}

We consider three server-side evasion methods that are able to forbid access to unwanted visitors: evasion via \file{.htaccess}, via \file{robots.txt}, via PHP. The \file{.htaccess} file allows personal configurations on a per-directory basis on Apache web servers. While a complete report on the functions of \file{.htaccess} is out of the scope of this paper, it is important to mention that this file can also serve as an access control filter~\cite{oest2018inside}. We consider a \file{.htaccess} file to be evasive if it contains \codeword{Forbidden} and \codeword{Redirection} rules that would prevent access to the phishing website. Similarly, the \file{robots.txt} file can be configured to deny search engine crawlers to scan a website. This is a desirable behavior for phishing kit authors, as these crawlers represent the primary component of some modern anti-phishing tools~\cite{khonji2013phishing, bell2020analysis}.
We define evasion via \file{robots.txt} if it contains a \codeword{Disallow} rule. 
The last evasion technique involves sophisticated methods implemented into the logical level of the application. We parse the Abstract Syntax Tree (AST) of each PHP file and consider evasion through PHP if we detect blacklist mechanisms or victim re-directions to some external pages.

We find a total of $920$ ($41\%$) phishing kits employing at least one evasion technique.
Table \ref{tab:evasion_rate} reports the numbers and rates of each evasion method observed in the considered kits. The rates referring to the different methods are expressed in terms of the total number of kits with evasion functions.
The \file{.htaccess} and PHP techniques seem the most common methods with their respective rates well above $50\%$. On the other hand, evasion through the \file{robots.txt} file seems slightly less preferred by the phishing kit authors, but it is still largely employed ($47\%$). The sum of percentages well above $100$\% confirms that when a phishing kit uses an evasion technique, it adopts more than one method to further decrease the probability of detection.

\begin{table}[ht]
    \centering
    \caption{Evasion techniques.}
    \label{tab:evasion_rate}
    \resizebox{0.80\columnwidth}{!}{
        \begin{tabular}{|c|c|c|}
            \hline
            \rowcolor[HTML]{EFEFEF} 
            \textbf{Category} & \textbf{Technique} & \textbf{Number (Rate)} \\ \hline \hline
            & \file{.htaccess} & $494$ (53\%)\\ \cline{2-3}
            & \file{robots.txt} & $433$ (47\%) \\ \cline{2-3}
            \multirow{-3}{*}{\rotatebox[origin=c]{90}{\textbf{Evasion}}} & PHP & $578$ (63\%) \\ \hline
        \end{tabular}
    }
\end{table}

\subsection{Obfuscation}

We identify five types of obfuscation techniques. Through the AST representation of the PHP code, we detect the illicit use of the \codeword{urldecode()} and \codeword{eval()} functions, which dynamically translate and evaluate an obfuscated string. Another popular obfuscation technique involves \codeword{base64} string encoding. Furthermore, we detect the illicit usage of hexadecimal assignments. While modern PHP interpreters automatically evaluate hexadecimal variable names at run time, this behavior can be exploited by authors to further confuse their source codes. Finally, we consider \textit{obfuscators}, which are automatic tools to obfuscate a portion of the source code. After a thorough study of several obfuscators, we verified that each of them presents distinguishable patterns that characterize their obfuscation method. From these patterns, we are able to define ad-hoc heuristics and regular expressions to detect their usage in the source code of the phishing kits.

Unlike extensive employment of evasion functions, we find that just $376$ ($17\%$) kits make use of obfuscation. A possible interpretation is that, unlike malware authors, phishing kit authors are more interested in decreasing the chances of detection through evasion (and thus increasing the time in which the phishing websites are up and running), rather than in hiding their source code through obfuscation. Similarly to the previous case, we report in Table~\ref{tab:obfuscation_rate} the specific rates for each obfuscation technique. The results denote that the \codeword{eval} function has been found in $77\%$ of the obfuscated kits, and clearly represents the favorite obfuscation method. Although few authors use automatic obfuscator tools (just 9\%), there is not a so clear preferred obfuscation function in the other techniques which present rates around $50\%$. As for the evasion case, the sum of percentages above $100\%$ suggests that multiple techniques are adopted when an author wants to obfuscate his code.

\begin{table}[ht]
    \centering
    \caption{Obfuscation techniques.}
    \label{tab:obfuscation_rate}
    \resizebox{0.85\columnwidth}{!}{
        \begin{tabular}{|c|c|c|}
            \hline
            \rowcolor[HTML]{EFEFEF} 
            \textbf{Category} & \textbf{Technique} & \textbf{Number (Rate)} \\ \hline \hline
            & \codeword{url\_decode} & $174$ (46\%) \\ \cline{2-3}
            & \codeword{eval} & $292$ (77\%) \\ \cline{2-3}
            & Hexadecimal assignment & $211$ (56\%) \\ \cline{2-3}
            & \codeword{base64} & $163$ (43\%) \\ \cline{2-3}
            \multirow{-5}{*}{\rotatebox[origin=c]{90}{\textbf{Obfuscation}}} & Obfuscator tool& $34$ (9\%) \\ \hline
        \end{tabular}
    }
\end{table}

\subsection{Author profiling}
\label{sec:authors}

During our exploratory analysis, we observed that a significant number of kits were signed by their authors denoting how marketing is important even in the dark world. Hence, we parsed the source code of the kits by looking for general keywords or small phrases that would represent a signature (e.g., coded by, developed by), and we manually filter the results in order to reduce false positives. Despite the na\"ive signature extraction methodology, we found $214$ different signatures for a total of $514$ signed kits (about 22\% of the total number of phishing kits in our dataset).

Table~\ref{tab:top_authors} reports the most frequent signatures along with their respective number of evasive and obfuscated kits. We note how few top authors produce even dozens of phishing kits, thus confirming the hypothesis of a narrow authors community. Another confirmation of the complex business relationships existing among the authors comes from the result that several kits have more than one signature. This supports the hypothesis that modern phishing kits can result from the more or less explicit collaboration of multiple authors, although we cannot exclude the use of multiple nicknames for the same author or for groups of them.

These data are even more interesting if associated with the results of our deterministic classification. Indeed, we can now build ``profiles'' of the main threat actors to keep track of their typical habits in terms of the evasion and obfuscation techniques they use to implement in their kits. As an example,  authors such as \textbf{xbalti} and \textbf{ex-robotos} make an extensive usage of specific evasion techniques ($42$/$42$ kits and $16$/$16$ evasive kits, respectively). Others, such as \textbf{venza} and \textbf{l33bo-phishers}, seem less interested in evasive functions ($2$/$23$ and $13$/$33$ evasive kits, respectively). On the other hand, from the low rates of obfuscated kits, we note how it is more challenging to find such extensive usage of obfuscation methods.
These findings let us conclude not only that a small group of authors design and develop the majority of phishing kits, but also that they tend to maintain similar coding structures.

\begin{table}[ht]
    \centering
    \caption{Top five detected signatures and respective number of evasive and obfuscated kits.}
    \label{tab:top_authors}
    \begin{tabular}{|c|c|c|c|c|}
        \hline
        \rowcolor[HTML]{EFEFEF} 
        \textbf{Signature} & \textbf{Number of kits} & \textbf{Evasive kits} & \textbf{Obfuscated kits}\\ \hline \hline
        xbalti                  & $42$ & $42$ ($100\%$) & $3$ ($7\%$) \\ \hline
        l33bo-phishers          & $33$ & $13$ ($39\%$)  & $3$ ($9\%$) \\ \hline
        venza                   & $23$ & $2$ ($8\%$)    & $0$ ($0\%$) \\ \hline
        medpage                 & $18$ & $9$ ($50\%$)   & $1$ ($5\%$) \\ \hline
        ex-robotos              & $16$ & $16$ ($100\%$) & $0$ ($0\%$) \\ \hline
    \end{tabular}
\end{table}

\subsection{Deterministic classification - lessons learned}
\label{sec:det_lessons}
As discussed in the previous sections, the deterministic classification analyzes the source code of the phishing kit leveraging static techniques. The results allow us to obtain an overview of the usage of the principal evasion and obfuscation functions in modern phishing kits, and to shed a light on the coding habits of the main threat actors. Evasion functions are extensively employed ($920$ evasive kits), while there is a scarce use of obfuscation ($376$ obfuscated kits).
Web-server providers can leverage this first deterministic classification to improve current early detection mechanisms. For example, they can implement filtering methods that exclude any source of code that includes such evasion or obfuscation functions. At the same time, they can prioritize the upcoming countermeasures according to the most employed malicious techniques.

A similar trend that prefers evasion over obfuscation can be evinced from the author profiles presented in Section~\ref{sec:authors}. From them, we not only note that few authors design and implement dozens of phishing kits, but also that they maintain their coding habits, thus confirming the re-utilization hypothesis. These results represent a first important step to performing more complex authorship attribution operations, for example assigning a novel kit to an author that typically uses the same functions. Moreover, threat intelligence analysts can benefit from our methodology to keep track of the possible evolution and changes in the behavior of the principal threat actors. 

Although static analysis guarantees a low number of false positives, it is also unable to keep pace with unknown functions introduced in the continuously evolving phishing kits. For this reason, we propose to employ ML classifiers that take advantage of the re-utilization of portions of phishing kits to provide a probabilistic classification.
As discussed in Section~\ref{sec:phaseIII}, the idea is that they can learn the typical structural patterns of evasive and obfuscated kits and use them to extend investigations on yet unseen kits.

\section{Classification based on Machine Learning Algorithms}
\label{sec:ml_results}

The application of machine learning algorithms for the classification of phishing kits is an original contribution of this paper. 
As discussed above, our expectation is that machine learning models are able to learn the common patterns of evasive and obfuscated kits, in order to provide a probabilistic classification that is able to flag even the kits employing unseen evasion and obfuscation functions.

We treat our learning problem as a binary classification task that groups the kits in terms of the evasion and obfuscation labels. For this reason, we choose and evaluate six classifiers that have obtained appreciable results for binary tasks in other cybersecurity contexts (e.g., \cite{xin2018machine, apruzzese2018effectiveness}).
All the classifiers, outlined below, were implemented in Python3 through the \textit{scikit-learn} toolkit. 

The \textit{Decision Tree} is a non-parametric supervised learning method that creates easy to interpret models based on decision rules that are inferred from the features of the samples.

We also consider two variants of its ensemble version \textit{Random Forest} (RF), which is often indicated as the best performance classification algorithm in multiple domains~\cite{apruzzese2020deep}. RF uses multiple decision trees and considers the results of each of them for the final classification. Each tree is built from a random subset of the samples, and each decision rule is generated considering only a subset of features. The idea is to exploit the variance of the resulting classifiers and to obtain a more robust ensemble prediction that outperforms any result based on one component. We evaluate two variants: \textit{Random Forest 10} (RF10) and \textit{Random Forest 100} (RF100). The former contains $10$ decision trees where decision rules are built on one-third of the original features. The latter contains $100$ decision trees, but its rules are built on the square root of the number of the original features.

The \textit{Support Vector Machines} (SVMs) algorithm constructs a high dimensional space, in which each dimension represents one feature. While each sample of the training set is projected onto that space, the model learns how to efficiently separate the samples belonging to different classes. We consider the linear version (\textit{Linear SVM}) in which the separation is performed by a linear function.

The \textit{Gaussian Processes} are other interesting supervised learning techniques providing a probabilistic classification. This type of algorithm differs from the previous methods as it associates each input sample with a set of probabilities for each class, instead of single predictions. This choice allows the computation of prediction confidence intervals that can assist the model debugging. 

\textit{Naive Bayes} is another algorithm that produces a probabilistic classification. Unlike the Gaussian approach, it makes the intentionally na\"{i}ve assumption that all the features are independent from each other. Given a sample, it computes the probabilities for each label, and predicts the class which obtains the highest score. Despite the simple initial assumptions, we consider this algorithm because it is extremely scalable and does not require large sets of data to achieve acceptable results.

\subsection{Feature set}
\label{sec:features}
The feature set for the considered classifiers is reported in Table \ref{tab:feature_set} in~\ref{sec_appendix:features}. It is designed to include several pieces of information about the phishing kit structure (e.g., number of files, directories), the presence of meaningful files and directories (e.g., \file{.htaccess}, \file{robots.txt}), the use of development frameworks (e.g, wordpress, laravel), authors' design patterns and choices (e.g., exfiltration method, PHP functions and superglobal variables). The idea is that the considered features reflect the design patterns of evasive and obfuscated kits in order to allow robust classification performance. This preliminary operation is performed by parsing the source code of each kit in our dataset, and leveraging ad-hoc heuristics and regular expressions to extract the features.

\subsection{Experimental scenarios}
\label{sec:scenarios}

We consider three main evaluation scenarios that aim to prove three desired properties of the proposed machine learning classifiers: (i) the capacity to classify evasive and obfuscated kits; (ii) their robustness when just a limited amount of samples is used for training; (iii) their ability to detect even those kits that employ novel evasion and obfuscation techniques that were not considered during training. The prediction of evasion and obfuscation techniques are treated as separate binary problems. Hence, in each scenario we train two classifiers: one to detect evasive kits, the other to detect obfuscated kits.

Scenario 1 (S1) represents the baseline use case to build the proposed probabilistic classification. Through this scenario, we aim to prove whether our method is able to predict the presence of known evasion and obfuscation techniques in the phishing kits from design patterns and structural information. We build the training set of the classifiers following the best practices in  literature~\cite{arp2020and}: 80\% of samples are used for training and 20\% for testing; the positive (i.e., evasive or obfuscated kits) and negative samples (i.e., not evasive nor obfuscated kits) are distributed equally in the training set.

Scenario 2 (S2) changes the training/test sets ratio to prove the robustness of the considered classifiers. We aim to verify whether the predictive performance remains high even with a limited amount of training samples. A positive result will further confirm the re-utilization hypothesis on which we base our method. In other words, it can suggest that the structure and design patterns of evasive and obfuscated kits are repeated, and that our classifiers do not require numerous examples to learn them. This experiment also tests the validity of our proposal in real-world contexts, in which the available set of kits represents just a minor fraction of the whole corpus of existing and future phishing kits. For these reasons, in this scenario, we use a challenging alternative where just 20\% of samples are used for training and 80\% for testing.

Scenario 3 (S3) aims to demonstrate that our method is able to detect evasive and obfuscated kits that employ novel techniques that were not included during training. In order to prove this property, we modify the labeling procedure of Phase II described in Section~\ref{sec:phaseII}. If in the original process we flagged a kit as evasive (obfuscated) because it employs at least one of the three (five) techniques detected by the deterministic approach, here we exclude a specific technique from the labeling procedure, and we flag the kit according to the remaining ones. Finally, we train the classifiers on this reduced version of the labels, and test them on the positive samples of the excluded technique. For example, we can consider only evasion via \file{.htaccess} and \file{robots.txt} for the label, and test the resulting classifiers against the phishing kits which employ evasion techniques through PHP. As a consequence, we obtain five and three different classifiers, respectively for each evasion and obfuscation function. A high detection rate in this scenario suggests that the resulting probabilistic classification would be able to catch also the new functions that phishing kit authors may introduce in their future kits.
\section{Probabilistic classification results}
\label{sec:ml_evaluation}

We present the results of the three scenarios by adopting the well-known metrics \textit{Precision}, \textit{Recall} (or \textit{Detection Rate}) and \textit{F1-score}.
The equations for Precision, Recall and F1-score are reported below, where TP, FP and FN denote true positives, false positives and false negatives, respectively.
\begin{align}
    P & = \frac{TP}{TP + FP} & R & = \frac{TP}{TP + FN}
\end{align}
\begin{align}
    F1\_score & = 2*\frac{P*R}{P+R}\textbf{}
\end{align}

Considering as a positive result an evasive (obfuscated) kit, the Precision measures the ability of a classifier not to label as positive a negative kit. On the other hand, Recall (Detection Rate) is the ability of a classifier to find all the evasive (obfuscated) kits. F1-score combines these two metrics into one indicator.
We remark that the lack of previous probabilistic classifications of evasive and obfuscated kits based on machine learning algorithms prevents a comparison of our results against those of related literature.

\subsection{Scenario 1}
\label{sec:results_scenario1}
The results related to the first scenario are reported in Table \ref{tab:scenario1}, where the numbers in bold denote the best results for the corresponding metric. Table \ref{tab:evasion_scenario1} refers to the classification of evasive kits. From it we can observe how all the classifiers achieve an F1-score above $0.9$, thus denoting their proficiency to a successful classification of evasive kits. In particular, \textit{Random Forest 100} offers the best performance with an F1-score of $0.964$. This result is in line with expectations as \textit{Random Forest} is recognized in literature as one of the best performing algorithms for classification tasks~\cite{apruzzese2018effectiveness}. As for obfuscated kits (Table \ref{tab:obfuscation_scenario1}), we report a slight drop in performance, but the classifiers reach Recall scores often superior to $0.9$. This suggests that they are still able to detect and classify most obfuscated kits. Besides, we are more interested in a low number of false negatives (i.e., high Recall) than in a low number of false positives (i.e., high Precision) because otherwise, some obfuscated kits would pass unnoticed.

As most of the classifiers considered in this scenario achieve an efficient probabilistic classification from structural information of phishing kits and authors' design patterns, we can confirm the quality of our feature set and the re-utilization hypothesis on which we base the entire approach. 

\begin{table}[ht]
    \centering
    \caption{Performance of ML classifiers in Scenario 1}
    \label{tab:scenario1}
    \begin{subtable}[htbp]{0.95\columnwidth}
        \resizebox{0.95\columnwidth}{!}{
            \begin{tabular}{|c|c|c|c|}
                \hline
                \rowcolor[HTML]{EFEFEF} 
                \textbf{Classifier} & \textbf{F1-score} & \textbf{Precision} & \textbf{Recall} \\ \hline \hline
                \MLalgorithm{Linear SVM} & $0.937$ & $0.940$ & $0.934$ \\ \hline
                \MLalgorithm{Gaussian Process} & $0.936$ & $0.934$ & $0.939$ \\ \hline
                \MLalgorithm{Decision Tree} & $0.944$ & $0.964$ & $0.925$ \\ \hline
                \MLalgorithm{Random Forest 10} & $0.961$ & $0.970$ & $0.952$ \\ \hline
                \MLalgorithm{Random Forest 100} & \bm{$0.964$} & \bm{$0.982$} & \bm{$0.953$} \\ \hline
                \MLalgorithm{Naive Bayes} & $0.927$ & $0.958$ & $0.900$ \\ \hline
                
            \end{tabular}
        }
        \caption{Performance against \textit{Evasive} phishing kits.}
        \label{tab:evasion_scenario1}
    \end{subtable}
    
    \begin{subtable}[t]{0.95\columnwidth}
        \resizebox{0.95\columnwidth}{!}{
            \begin{tabular}{|c|c|c|c|}
                \hline
                \rowcolor[HTML]{EFEFEF} 
                \textbf{Classifier} & \textbf{F1-score} & \textbf{Precision} & \textbf{Recall} \\ \hline \hline
                \MLalgorithm{Linear SVM} & $0.745$ & $0.623$ & $0.927$ \\ \hline
                \MLalgorithm{Gaussian Process} & $0.723$ & $0.622$ & $0.870$ \\ \hline
                \MLalgorithm{Decision Tree} & $0.766$ & $0.655$ & $0.923$ \\ \hline
                \MLalgorithm{Random Forest 10} & $0.820$ & $0.754$ & $0.900$ \\ \hline
                \MLalgorithm{Random Forest 100} & \bm{$0.852$} & \bm{$0.787$} & \bm{$0.930$} \\ \hline
                \MLalgorithm{Naive Bayes} & $0.765$ & $0.667$ & $0.898$ \\ \hline
                
            \end{tabular}
        }
        \caption{Performance against \textit{Obfuscated} phishing kits.}
        \label{tab:obfuscation_scenario1}
    \end{subtable}
\end{table}

\subsection{Scenario 2}
\label{sec:results_scenario2}
The second scenario aims to test the robustness of the probabilistic classification by reducing the dimension of the training set to just the $20$\% of the kits. The results are reported in Table \ref{tab:scenario2}. Interestingly, we observe no significant performance drop with respect to the previous scenario. For example, the F1-scores of the classifiers for the evasion remain above $0.9$ or close to it. As before, we note lower Precision scores for the classifiers trained to detect obfuscation techniques. However, the high Recall scores indicate the ability to detect the majority of obfuscated kits.

We can conclude that the resulting classifiers are almost insensitive to the size of the training set. This robustness is an extremely important factor for the classification of future kits, and indicates that it is possible to obtain an effective probabilistic classification without requiring a large number of samples. Furthermore, we can also interpret these results as additional proof of the re-utilization hypothesis. Hence, we can affirm that the evasive and obfuscated kits preserve their characteristics, and that the structures and the design patterns of the kits are repeated. This phenomenon can be exploited by threat intelligence experts to carry out author profiles for specific threat actors, and to keep track of the evolution of their behavior through time.

\begin{table}[ht]
    \centering
    \caption{Performance of ML classifiers in Scenario 2}
    \label{tab:scenario2}
    \begin{subtable}[htbp]{0.95\columnwidth}
        \resizebox{0.90\columnwidth}{!}{
            \begin{tabular}{|c|c|c|c|}
                \hline
                \rowcolor[HTML]{EFEFEF} 
                \textbf{Classifier} & \textbf{F1-score} & \textbf{Precision} & \textbf{Recall} \\ \hline \hline
                \MLalgorithm{Linear SVM} & $0.882$ & $0.848$ & $0.923$ \\ \hline
                \MLalgorithm{Gaussian Process} & $0.896$ & $0.889$ & $0.904$ \\ \hline
                \MLalgorithm{Decision Tree} & $0.917$ & $0.913$ & $0.919$ \\ \hline
                \MLalgorithm{Random Forest 10} & $0.921$ & $0.931$ & $0.917$ \\ \hline
                \MLalgorithm{Random Forest 100} & \bm{$0.947$} & \bm{$0.962$} & \bm{$0.934$} \\ \hline
                \MLalgorithm{Naive Bayes} & $0.895$ & $0.923$ & $0.869$ \\ \hline
                
            \end{tabular}
        }
        \caption{Performance against \textit{Evasive} phishing kits.}
        \label{tab:evasion_scenario2}
    \end{subtable}
    
    \begin{subtable}[t]{0.95\columnwidth}
        \resizebox{0.90\columnwidth}{!}{
            \begin{tabular}{|c|c|c|c|}
                \hline
                \rowcolor[HTML]{EFEFEF} 
                \textbf{Classifier} & \textbf{F1-score} & \textbf{Precision} & \textbf{Recall} \\ \hline \hline
                \MLalgorithm{Linear SVM} & $0.725$ & $0.602$ & $0.911$ \\ \hline
                \MLalgorithm{Gaussian Process} & $0.661$ & $0.523$ & $0.901$ \\ \hline
                \MLalgorithm{Decision Tree} & $0.744$ & $0.732$ & $0.758$ \\ \hline
                \MLalgorithm{Random Forest 10} & $0.773$ & $0.686$ & $0.890$ \\ \hline
                \MLalgorithm{Random Forest 100} & \bm{$0.833$} & \bm{$0.758$} & \bm{$0.926$} \\ \hline
                \MLalgorithm{Naive Bayes} & $0.730$ & $0.606$ & $0.917$ \\ \hline
                
            \end{tabular}
        }
        \caption{Performance against \textit{Obfuscated} phishing kits.}
        \label{tab:obfuscation_scenario2}
    \end{subtable}
\end{table}

\subsection{Scenario 3}
\label{sec:results_scenario3}
The third scenario tests the ability of the classifiers to detect novel evasion and obfuscation techniques that were not included during training. For a complete analysis, we carry out several classifiers. As discussed in Section~\ref{sec:scenarios}, each classifier is trained respectively on a different version of the training set labeled excluding one of the evasion (obfuscation) functions from the labeling procedure. They are then tested against the kits which use the specific excluded technique to assess whether they can still flag them as evasive (obfuscated). For this reason, the results are reported in terms of Detection Rate. 

If we consider evasion (Table \ref{tab:evasion_scenario3}), we can observe that the algorithms are capable of correctly classifying as evasive the majority of kits using an unseen method. The results show an average Detection Rate of over $84$\%. We can observe similar Detection Rates also for obfuscation techniques (Table \ref{tab:obfuscation_scenario3}) with \textit{Random Forest 10} and \textit{Random Forest 100} classifiers offering the best results. We highlight that some algorithms (e.g., \textit{Linear SVM} and \textit{Gaussian Process}) present difficulties in detecting some specific techniques. As this can be caused by the limited number of kits in the training sets used for those techniques, it also denotes that these methods present margins of improvements.

\begin{table}[h]
    \centering
    \caption{Detection Rate for Evasion Techniques in Scenario 3.}
    \label{tab:evasion_scenario3}
    \begin{tabular}{|c|c|c|c|}
        \hline
        \rowcolor[HTML]{EFEFEF} 
        & \multicolumn{3}{c|}{\textbf{Predicted Technique}} \\ \cline{2-4}
        \rowcolor[HTML]{EFEFEF} 
         \multirow{-2}{*}{Classifier} & \file{.htaccess} & \file{robots.txt} & PHP \\ \hline \hline
         \MLalgorithm{Linear SVM} & $71$\% & $89$\% & $86$\% \\ \hline
         \MLalgorithm{Gaussian Process} & $77$\% & $87$\% & $83$\% \\ \hline
         \MLalgorithm{Decision Tree} & $74$\% & $83$\% & $79$\% \\ \hline
         \MLalgorithm{Random Forest 10} & $77$\% & $87$\% & $89$\% \\ \hline
         \MLalgorithm{Random Forest 100} & $73$\% & $92$\% & $96$\% \\ \hline
         \MLalgorithm{Naive Bayes} & \bm{$81$\%} & \bm{$100$\%} & \bm{$98$\%} \\ \hline
        
    \end{tabular}
\end{table}

\begin{table}[h]
    \centering
    \caption{Detection Rate for Obfuscation Techniques in Scenario 3.}
    \label{tab:obfuscation_scenario3}
    \resizebox{0.95\columnwidth}{!}{
        \begin{tabular}{|c|c|c|c|c|c|}
            \hline
            \rowcolor[HTML]{EFEFEF} 
            & \multicolumn{5}{c|}{\textbf{Predicted Technique}} \\ \cline{2-6}
            \rowcolor[HTML]{EFEFEF} 
             \multirow{-2}{*}{Classifier} & \codeword{url\_decode} & \codeword{eval} & \codeword{hex} & \codeword{base64} & \codeword{obfuscator} \\ \hline \hline
             \MLalgorithm{Linear SVM} & $79$\% & $51$\% & $70$\% & $70$\% & $32$\% \\ \hline
             \MLalgorithm{Gaussian Process} & $80$\% & $48$\% & $70$\% & $65$\% & $35$\%\\ \hline
             \MLalgorithm{Decision Tree} & $77$\% & $65$\% & $76$\% & $84$\% & $76$\%\\ \hline
             \MLalgorithm{Random Forest 10} & $87$\% & \bm{$84$\%} & \bm{$91$\%} & $85$\% & $74$\%\\ \hline
             \MLalgorithm{Random Forest 100} & \bm{$89$\%} & $83$\% & $90$\% & \bm{$85$\%} & \bm{$77$\%}\\ \hline
             \MLalgorithm{Naive Bayes} & $83$\% & $49$\% & $80$\% & $71$\% & $35$\% \\ \hline
            
        \end{tabular}
    }
\end{table}

\subsection{Probabilistic classification - lessons learned}
\label{sec:prob_lessons}

Probabilistic classification has been introduced to overcome the limitations of the deterministic phase that is unable to keep pace with novel techniques. We investigate whether structural and design patterns of phishing kits can be leveraged by ML classifiers to correctly flag evasive and obfuscated kits. We can conclude that this approach based on machine learning can surpass the issues of an exclusively static approach that has been typically used in related work.
Moreover, our experimental campaign proves that the proposed methodology leads to robust classifiers which also detect \textbf{novel} evasion and obfuscation methods. This property acquires great importance in real world scenarios as it provides web-server providers with an instrument capable of flagging possible unknown malicious functions which can thus be rapidly identified to devise more effective countermeasures. 
\section{Related work}
\label{sec:related}

This paper presents the first deterministic and probabilistic classification based on a large set of phishing kits. The proposed method is able to group kits according to their evasion and obfuscation functions and to detect some design patterns adopted by the authors.
The literature considers detailed analyses of specific phishing kit functions and techniques or adopts different methods that are unable to get conclusions similar to those achieved in this paper.

The authors in \cite{oest2018inside} analyze and classify the fake URLs that attackers use to deceive the victims of phishing and the \textit{.htaccess} server-side filtering techniques to evade detection. From this analysis, they conclude that phishers tend to reuse parts of the same phishing kit in multiple campaigns. We can confirm this interesting hypothesis by considering multiple features of phishing kits besides specific URLs. The same conclusions are drawn by the authors in~\cite{merlo2022phishing} who analyze the source code similarity distribution among phishing kits. We leverage this thesis to train our machine learning classifiers that are able to detect even novel evasion and obfuscation techniques because they are often combinations or modifications of existing components. Moreover, we further demonstrate that the main authors tend to maintain the same coding habits over time.

Other surveys of phishing attacks consider attack types, communication media, vectors and target devices in their analysis (e.g., \cite{aleroud2017phishing, chiew2018survey}), but they are far from the in-depth characterization of phishing kits proposed in this paper. A compelling taxonomy based on phishing environments and anti-phishing methods is proposed in \cite{aleroud2017phishing}, but even this is focused on the phishing countermeasures that are out of the scope of this paper. Moreover, its analysis does not deal with phishing kits nor with evasion and obfuscation techniques. Similar limitations affect also the interesting results presented in \cite{chiew2018survey}.

Other authors leverage a Web honeypot to attract attackers into deploying their kits on a controlled application~\cite{han2016phisheye}. This honeypot allows the researchers to collect over 600 kits and to estimate an average lifetime of eight days before anti-phishing detection. This paper presents an original white-box analysis of the phishing kits' behavior in live environments. Although its goal is not oriented to classify the kits on the basis of design patterns nor of evasion and obfuscation techniques, this work represents an interesting complementary alternative to our approach to extract dynamic features from real-world phishing kits.

Extensive research (e.g., \cite{cova2008there, oest2020sunrise, ho2019detecting}) has been oriented to analyze single kit functions, to monitor their lifecycle or is mainly interested in detection goals. The results are interesting but they are not intended to provide a general classification of multiple phishing kits. For example, the authors in~\cite{cova2008there} study the structural design and implementation of free and crawled phishing kits with the main goal of identifying the obfuscation procedures to hide the backdoors for exfiltrating information. Our classification considers several obfuscation techniques that do not explicitly relate to backdoors.
An analysis of \emph{lateral phishing} attacks is proposed in~\cite{ho2019detecting}. However, the authors are not interested in extending their discussions to other evasion or obfuscation techniques. 
The paper in \cite{oest2020sunrise} presents Golden Hour which is a framework for the proactive detection and mitigation of phishing attacks. The proposed method relies on the observation that modern phishing pages make substantial requests for resources to the Web servers of the target organization. By correlating data from multiple sources, the authors are able to propose a novel approach for phishing detection.

Machine learning techniques represent a novelty of this paper, hence we consider important to evidence our original contribution with respect to this research area.
Supervised machine learning techniques are typically related to the detection of phishing attacks (e.g., \cite{abdelhamid2017phishing, el2020depth}). Several examples adopt machine learning algorithms for fast detection mechanisms that are based on features extracted from phishing URLs and HTML pages (e.g.,~\cite{blum2010lexical, xiang2011cantina+, chiew2019new, sonmez2018phishing, ho2019detecting, sahingoz2019machine, peng2018detecting, rao2019detection, jain2019machine, wu2019phishing}). 
The feature set proposed in this paper consists of multiple detected design patterns.
For our classification purposes, design patterns represent a more efficient alternative with respect to methods based just on URLs and HTML page features. This is because considering the coding habits of the authors allows us to also keep track of their evolution over time and to assist future attribution operations.
Even more importantly, we leverage machine learning algorithms not to detect phishing attacks, but as an original way to support phishing kit classification and even to predict the possibility of novel evasion and obfuscation techniques adopted by the phishing kit authors. For these reasons, our methods and results are far from unsupervised approaches that are oriented to detect and cluster phishing campaigns and to cluster phishing Web pages for detection and attribution purposes (e.g.,~\cite{layton2010automatically, liu2010automatic, layton2012unsupervised, britt2012clustering, cui2017tracking}).

\section{Conclusions}
\label{sec:conclusions}
Phishing kits are ready-to-use tools that allow a rapid deployment of a phishing website. They are distributed in the dark market and represent a serious menace to cybersecurity as they grant the possibility to carry out serious phishing attacks even to unskilled criminals. While previous works focus on specific kit functions or are primarily interested in detection goals, we present a novel classification based on more than 2000 recent phishing kits.
Our classification method is based on two complementary approaches. The deterministic classification performs a static analysis of the source code of the phishing kits, and extracts information about the adopted evasion and obfuscation techniques. We verify that evasion techniques are largely more employed than obfuscation ones. Moreover, we provide a first example of author profiling. By grouping the kits according to the presented signatures, we also demonstrate that the authors tend to maintain the same coding habits through time. Despite the low number of false positives, this first deterministic phase is unable to detect novel functions that the authors may introduce. 
For this reason, we also propose the adoption of ML classifiers that are trained to detect evasive and obfuscated kits from repeated structures and design patterns. The hypothesis is that the small group of the most prolific authors tend not to change their programming habits and reuse parts of previous kits in novel products. We confirm this hypothesis through multiple scenarios that prove that the proposed machine learning methods lead to robust classifiers which learn the typical patterns of the majority of evasive and obfuscated kits, do not require a large number of examples in the training set, and are also capable of correctly classifying kits that employ novel techniques that were unseen during training. 
This paper should be considered as an initial contribution to the analysis of phishing kits that represent a consistent problem of modern cybercrime. Future work can leverage the achieved results in a twofold direction: the improvement of detection tools through the analysis of the considered evasion and obfuscation techniques; the design and evaluation of more advanced threat intelligence operations based on author profiles and their adopted schemes. We evidence also the possibility of integrating our results with stylometry techniques that are widely used for the attribution of literary texts to their authors. 

\appendix
\newpage
\section{Feature set}
\label{sec_appendix:features}

\setlength\LTleft{-65pt}
\begin{longtable}{|m{0.7cm}|c|c|c|}
    \caption{Feature set for the considered classifiers}
    \label{tab:feature_set}
    \\ \hline
    \rowcolor[HTML]{EFEFEF} 
    \textbf{} & \textbf{Feature} & \textbf{Description} & \textbf{Type} \\ \hline \hline
    \endhead
     & \textit{nFiles} & Number of files & Numerical \\ \cline{2-4}
     & \textit{nDir} & Number of directories & Numerical\\ \cline{2-4}
     & \textit{nPhp} & Number of PHP files & Numerical \\ \cline{2-4}
     & \textit{nJs} & Number of Javascript files & Numerical \\ \cline{2-4}
     & \textit{nTxt} & Number of text files & Numerical \\ \cline{2-4}
     & \textit{nExe} & Number of Executables & Numerical\\ \cline{2-4}
     & \textit{nDll} & Number of dynamic-link libraries & Numerical \\ \cline{2-4}
     & \textit{nApk} & Number of Android Package files & Numerical \\ \cline{2-4}
     & \textit{nHtml} & Number of HTML files & Numerical \\ \cline{2-4}
     & \textit{nCss} & Number of CSS files & Numerical \\ \cline{2-4}
     & \textit{nPdf} & Number of PDF files & Numerical \\ \cline{2-4}
     & \textit{nMul} & Number of multimedia (images and videos) files & Numerical \\ \cline{2-4}
    \hspace{3pt}\multirow{-13}{*}{\rotatebox[origin=c]{90}{\textbf{Structural features}}} & \textit{Otherfiles} & Number of other files & Numerical \\ \hline \hline
    
     & \file{htaccess} & Presence of \file{.htaccess} file & Bool \\ \cline{2-4}
     & \file{robots.txt} & Presence of \file{robots.txt} file & Bool \\ \cline{2-4}
     & \textit{admin} & Presence of files or directories named "admin" & Bool \\ \cline{2-4}
    \hspace{5pt}\multirow{-4}{*}{\rotatebox[origin=c]{90}{\multirow{2}{*}{\textbf{\begin{tabular}[c]{@{}c@{}}Relevant \\ files \end{tabular}}}}} & \textit{config} & Presence of files or directories which contains "config" in name & Bool \\ \hline \hline
    
     & \textit{wordpress} & Use of Wordpress & Bool \\ \cline{2-4}
     & \textit{laravel} & Use of Laravel & Bool \\ \cline{2-4}
     & \textit{code\_ign} & Use of CodeIgniter & Bool \\ \cline{2-4}
     & \textit{zend} & Use of Zend & Bool \\ \cline{2-4}
    \hspace{4pt}\multirow{-5}{*}{\rotatebox[origin=c]{90}{\multirow{2}{*}{\textbf{\begin{tabular}[c]{@{}c@{}}Development \\ Frameworks \end{tabular}}}}} & \textit{HTTrack} & Use of website copier HTTrack & Bool \\ \hline \hline

    \newpage \hline
     & \textit{api\_call} & Call to external API & Bool \\ \cline{2-4}
     & \textit{array\_hostnames} & Presence of array of hostnames & Bool \\ \cline{2-4}
     & \textit{array\_ipaddresses} & Presence of array of IP addresses & Bool \\ \cline{2-4}
     & \textit{form\_validation} & Form validation & Bool \\ \cline{2-4}
     & \textit{deceiving\_url} & Similarity between deployment url and target organization name & Bool \\ \cline{2-4}
     & \textit{deceiving\_zipname} & Similarity between zip name and target organization name & Bool \\ \cline{2-4}
     & \textit{read\_file} & Presence of \codeword{read} function call & Bool \\ \cline{2-4}
     & \textit{redirection} & Presence of redirection function & Bool \\ \cline{2-4}
     & \textit{random\_file} & Creation of randomly named files & Bool \\ \cline{2-4}
    \hspace{4pt}\multirow{-11}{*}{\rotatebox[origin=c]{90}{\multirow{2}{*}{\textbf{\begin{tabular}[c]{@{}c@{}}PHP Functions \end{tabular}}}}} & \textit{random\_dir} & Creation of randomly named directories & Bool \\ \hline \hline
    
     & \codeword{\$\_SERVER} & Use of \codeword{\$\_SERVER} superglobal variable & Bool \\ \cline{2-4}
     & \codeword{\$\_GET} & Use of \codeword{\$\_GET} superglobal variable & Bool \\ \cline{2-4}
     & \codeword{\$\_POST} & Use of \codeword{\$\_POST} superglobal variable & Bool \\ \cline{2-4}
     & \codeword{\$\_FILES} & Use of \codeword{\$\_FILES} superglobal variable & Bool \\ \cline{2-4}
     & \codeword{\$\_REQUEST} & Use of \codeword{\$\_REQUEST} superglobal variable & Bool \\ \cline{2-4}
     & \codeword{\$\_SESSION} & Use of \codeword{\$\_SESSION} superglobal variable & Bool \\ \cline{2-4}
     & \codeword{\$\_ENV} & Use of \codeword{\$\_ENV} superglobal variable & Bool \\ \cline{2-4}
    \hspace{4pt}\multirow{-8}{*}{\rotatebox[origin=c]{90}{\multirow{2}{*}{\textbf{\begin{tabular}[c]{@{}c@{}}Supergobal variables \end{tabular}}}}} & \codeword{\$\_COOKIE} & Use of \codeword{\$\_COOKIE} superglobal variable & Bool \\ \hline \hline
    
     & \textit{mail} & Presence of \codeword{mail} function call & Bool \\ \cline{2-4}
     & \textit{bot\_telegram} & Presence of Telegram Bot API & Bool \\ \cline{2-4}
    \hspace{4pt}\multirow{-3}{*}{\rotatebox[origin=c]{90}{\multirow{2}{*}{\textbf{\begin{tabular}[c]{@{}c@{}}Exf.\\ method\end{tabular}}}}} & \textit{write} & Presence of \codeword{write} function call & Bool \\ \hline
    
\end{longtable}

\newpage
\bibliographystyle{plain}
\bibliography{bibliography}


\end{document}